\documentstyle[twoside,fleqn,espcrc2,epsfig]{article}

\newcommand{\etal}{{\it et al.}}
\title{Quark Loop Effects with an Improved Staggered Fermion Action
\thanks{Presented by C.~DeTar.}}
\author{ C.~Bernard
\address{Department of Physics, Washington University, St.~Louis, MO 63130, USA},
T.~Burch
\address{Department of Physics, University of Arizona, Tucson, AZ 85721, USA}, 
T.A.~DeGrand
\address{Physics Department, University of Colorado, Boulder, CO 80309, USA},
C.E.~DeTar
\address{Physics Department, University of Utah, Salt Lake City, UT
  84112, USA},
Steven~Gottlieb
\address{Department of Physics, Indiana University, Bloomington, IN 47405, USA},
U.M.~Heller
\address{CSIT, Florida State University, Tallahassee, FL 32306-4120, USA},
K.~Orginos$\,\null^{\rm b}$,
R.L.~Sugar
\address{Department of Physics, University of California, Santa Barbara, CA 93106, USA},
and D.~Toussaint$\,\null^{\rm b}$,
} 
\begin{document}

\begin{abstract}
We have been studying effects of dynamical quarks on various hadronic
observables, using our recently formulated improvement for staggered
fermions.  To illustrate improvement, we show that the light hadron
spectrum in the quenched approximation gives remarkably good scaling.
We highlight three new results: (1) We find no apparent quark
loop effects in the Edinburgh plot with 2+1 flavors of dynamical
quarks at $a = 0.14$ fm. (2) We show that dynamical quarks modify the
shape of the heavy quark potential. (3) We present results hinting at
meson decay effects in light hadron spectroscopy.
\end{abstract}

\maketitle 
\section{INTRODUCTION}

Recent proposals for Symanzik improvement of the staggered fermion
action have produced encouraging results, particularly in an improved
flavor and rotational symmetry \cite{MILC_FATTEST,Orginos:2000kg}.
Adding a few terms to the conventional Kogut-Susskind action, namely
three-link, five-link and seven-link staples and a third-neighbor
coupling, removes all tree-level ${\cal O}(a^2)$ errors
\cite{NAIK,MILC_FAT,MILC_NAIK,ILLINOIS_FAT,MILC_FATTER,LEPAGE_TSUKUBA,LEPAGE98}.
The computational price for using this action is quark mass dependent,
and is roughly a factor of 2.5 for the lightest masses we
are using.
Here we extend previous work with our
preferred ``Asqtad'' action, studying scaling in the quenched
approximation, and exploring the effects of dynamical quarks on the
light hadron spectrum and the heavy-quark potential.

We have accumulated a library of gauge configurations, both quenched
(one-loop Symanzik improved gauge action) and
dynamical.\footnote[1]{The MILC code, including the improved KS action code, is
publicly available.  Gauge configurations are also available.  Please
contact doug@physics.arizona.edu or detar@physics.utah.edu for
details}
Our
\begin{center}
\setlength{\tabcolsep}{1.5mm}
\begin{tabular}{|l|l|c|l|l|}
\hline
$m_{u,d}$ / $m_s$  & \hspace{-1.0mm}$10/g^2$  & size            & lats. & $a/r_1$ \\
\hline                             
                       & 8.40  & $28^3\!\times\! 96$ &  101$^*$ & 0.2683(9) \\
                       & 8.00  & $20^3\!\times\! 64$ &  408 & 0.3753(8) \\
 	     quenched  & 7.75  & $16^3\!\times\! 32$ &  206 & \\
                       & 7.60  & "               &  100 & \\
                       & 7.40  & "               &  191 & \\
\hline                             
0.40      / 0.40   &     7.35  & $20^3\!\times\! 64$ &  335 & 0.3757(10) \\
0.20      / 0.20   &     7.15  & "               &  349 & 0.3700(10) \\
0.10      / 0.10   &     6.96  & "               &  344 & 0.3721(13) \\
0.05      / 0.05   &     6.85  & "               &  421 & 0.3732(15) \\
0.04      / 0.05   &     6.83  & "               &  208$^*$ & 0.3751(19) \\
0.03      / 0.05   &     6.81  & "               &  474$^*$ & 0.3749(14) \\
0.02      / 0.05   &     6.79  & "               &  368$^*$ & 0.3762(13) \\
0.01      / 0.05   &     6.76  & "               &  128$^*$ & 0.3848(25) \\   
\hline
\end{tabular}
\end{center}
\vspace{-3.0mm}
$^*$ sample is currently being enlarged.
\vspace{2.0mm}

\noindent
dynamical fermion lattices are generated in the
presence of two
lighter flavors ($u$, $d$) and one heavier
flavor ($s$).
For the dynamical fermion sample 
the gauge coupling is tuned so that the lattice spacing
remains the same as the quark mass is varied, allowing an exploration
of quark mass effects independent of scale.  The current parameter set
is shown in the table above.

\section{SCALING TESTS}
\begin{figure}[t]
 \vspace*{-5mm} 
\hspace{10mm}
 \epsfig{bbllx=0,bblly=0,bburx=4000,bbury=4000,clip=,
 file=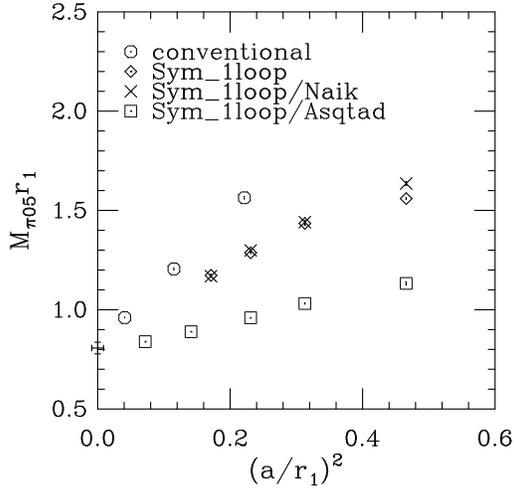,width=74mm} \vspace*{-5mm}
\caption{Flavor symmetry test.  Mass of the non-Goldstone local pion
{\it vs} lattice spacing in units of $r_1$ (see text) for progressively
improved actions: (octagon) conventional staggered fermion and
single-plaquette gauge action, (diamond) same but with an improved
gauge action, (cross) same but with the Naik third-neighbor term, and
(square) our preferred Asqtad action.
\label{fig:pi05_vs_m_2}
}
\vspace*{-7mm}
\end{figure}
\begin{figure}[t]
 \vspace*{-5mm}
 \epsfig{bbllx=0,bblly=0,bburx=4000,bbury=4000,clip=,
 file=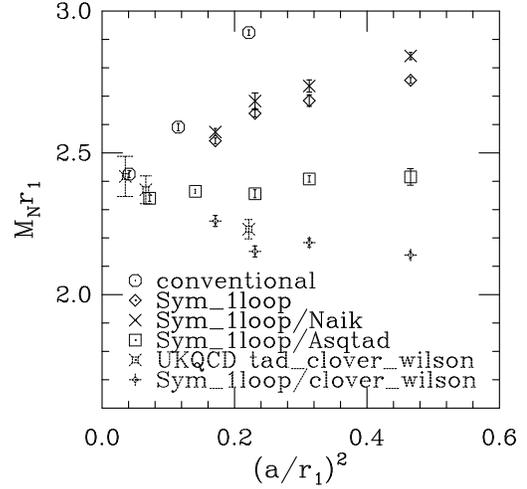,width=74mm} \vspace*{-5mm}
\caption{Scaling test.  Nucleon mass {\it vs} lattice spacing.  Same
as \protect\ref{fig:pi05_vs_m_2}, but including Wilson-clover results.
\label{fig:nuc_vs_m_2}
}
\vspace*{-7mm}
\end{figure}
\begin{figure}[t]
 \vspace*{-5mm}
 \epsfig{bbllx=0,bblly=0,bburx=4000,bbury=4000,clip=,
 file=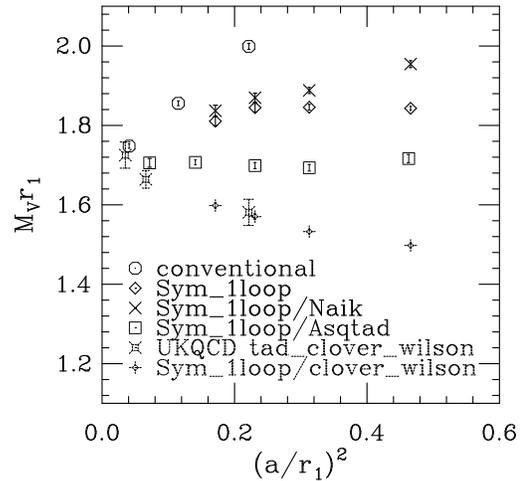,width=74mm}  \vspace*{-5mm}
\caption{Same as \protect\ref{fig:nuc_vs_m_2}, but
for the rho meson mass.
\label{fig:rho_vs_m_2}
}
\vspace*{-5mm}
\end{figure}

To measure the degree of residual flavor symmetry breaking, we compute
the mass of the local non-Goldstone pion ($\pi_{05}$) as a function of
lattice spacing at a fixed value of the Goldstone pion mass ($m_G$) on
a series of quenched lattices \cite{IMP_SCALING}.  In particular, we
fix $m_G r_1 = 0.807(3)$ (Goldstone $\pi$) with the scale $r_1$ set by
the force $F_{\rm Q\bar Q static}$ between static quarks $r_1^2 F_{\rm
Q\bar Q static}(r_1) = 1$, a variant of Sommer's
scale\cite{SOMMER_R0}. 
The value of $r_0/r_1$ depends on the quark masses, varying from
1.376(2) for the quenched theory to 1.44(1) for physical quark masses.
The effect of
improving the fermion action is shown in
Fig.~\ref{fig:pi05_vs_m_2}.  It is expected that the curves for all
actions extrapolate to the Goldstone pion mass.

Scaling of the $\rho$ and nucleon masses is shown in
Figs.~\ref{fig:nuc_vs_m_2} and \ref{fig:rho_vs_m_2}.  Results for
variants of the Wilson-clover action are also shown
\cite{SCRIWILSON,UKQCDWILSON}.  For these masses, the Asqtad action
gives the best result.  Nonetheless, it is encouraging that staggered
fermion and Wilson-clover simulations agree
in the continuum limit.

\section{DYNAMICAL QUARK EFFECTS: EDINBURGH PLOT}

\begin{figure}
 \vspace*{-5mm} \epsfig{bbllx=0,bblly=0,bburx=4000,bbury=4000,clip=,
 file=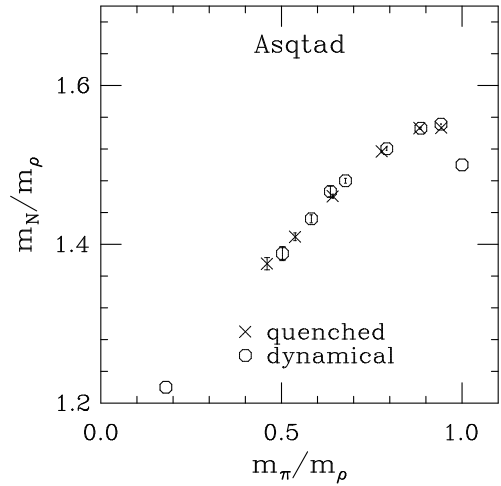,width=74mm} \vspace*{-8mm}
\caption{Edinburgh plot comparing results for a quenched and $2 + 1$
flavor dynamical simulation at $a = 0.14$ fm.  Octagons show the
experimental value (lower left) and infinite quark mass value (upper
right).
\label{fig:edin_both}
}
\vspace*{-7mm}
\end{figure}

To determine the effect of dynamical quarks on the light hadron
spectrum, we ran a series of quenched and $2+1$ dynamical fermion
simulations with gauge couplings tuned to fix the lattice spacing $a =
0.14$ fm, using $20^3 \times 64$ lattices for both the quenched and
dynamical runs.  The plotted points are from the runs described in
Table I with light quark masses down to $0.02 a^{-1}$ and from the
quenched run at $10/g^2=8.0$.  The resulting Edinburgh plot in
Fig.~\ref{fig:edin_both} shows no discernible change when dynamical
quarks are introduced.  This $2+1$ flavor $a = 0.14$ fm result appears
to be in conflict with our previous claims, based on the conventional
action with two sea quark flavors in the $a \rightarrow 0$ limit
\cite{CONVENTIONALKS}.  We are currently investigating whether the
difference comes from the number of flavors, the continuum
extrapolation, or some other source of systematic error.

\section{DYNAMICAL QUARK EFFECTS: HEAVY QUARK POTENTIAL}
\begin{figure}
 \vspace*{-5mm}
 \epsfig{bbllx=0,bblly=0,bburx=4000,bbury=4000,clip=,
 file=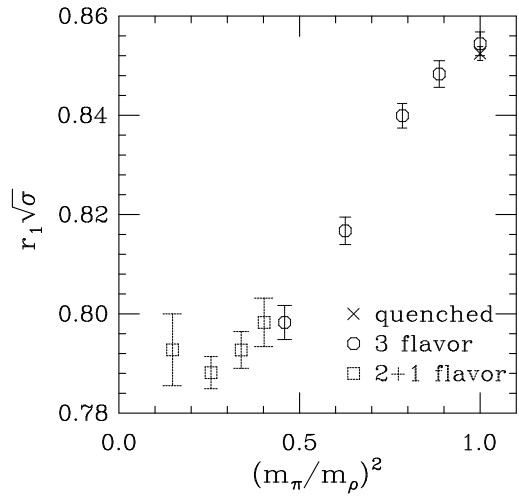,width=74mm}
 \vspace*{-8mm}
\caption{Square root of the string tension in units of $r_1$
{\it vs} the squared ratio of the pion to rho mass.
The octagons are done with three degenerate flavors (for $m_q \ge m_s$),
while the squares are runs with two light flavors and a fixed strange
quark mass.
\label{fig:r1_sq_sigma} 
}
\vspace*{-7mm}
\end{figure}
\begin{figure*}[tb]
 \vspace*{-10mm} \hspace{10mm}
 \epsfig{bbllx=0,bblly=0,bburx=4000,bbury=4000,clip=,
 file=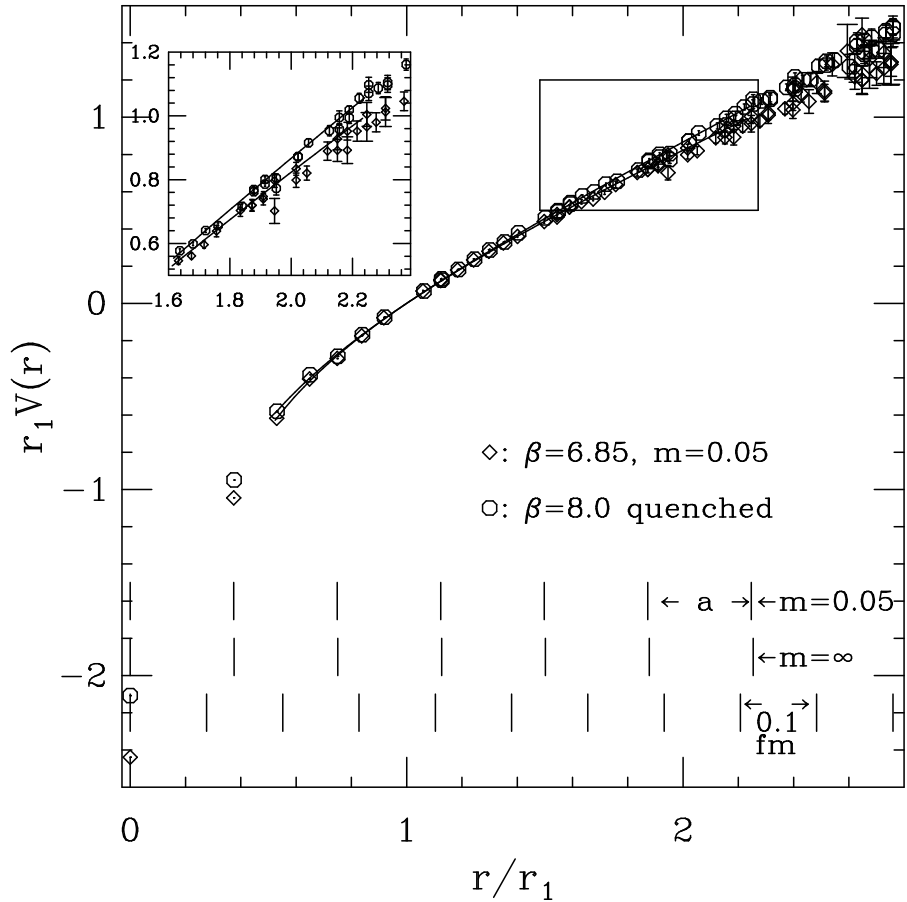,width=130mm}
 \vspace*{-10mm}
\caption{Heavy quark potential with and without $2+1$ flavors of sea
quarks.  Rulers show lattice and physical units for the two matched
simulations.
\label{fig:potmatch_phys.b800b685l20} 
}
\vspace*{-2mm}
\end{figure*}

Dynamical quark loops modify the heavy quark potential
\cite{Bernard:2000gd}, {\it e.g.}, by opening the two-meson decay
channel and by altering the running of the Coulomb coupling.  In
Fig.~\ref{fig:potmatch_phys.b800b685l20} we plot the heavy quark
potential with and without $3$ flavors of sea quarks of mass $am_q =
0.05$.  The vertical scale is adjusted to give agreement at
$r_1$, and since the force at $r_1$ is used to set the scale, the
two potentials have the same slope there.

To the extent the potential changes shape under the influence of sea
quarks, setting the lattice scale from the heavy quark potential
clearly has limitations.  This point is illustrated in
Fig.~\ref{fig:r1_sq_sigma}, where we plot the ratio of two possible
heavy-quark-potential-inspired scales as a function of the ratio of
the pion to rho mass.

\section{DYNAMICAL QUARK EFFECTS: MESON DECAY}
\begin{figure}[t]
 \vspace*{-11mm}
 \epsfig{bbllx=0,bblly=0,bburx=4000,bbury=4000,clip=,
 file=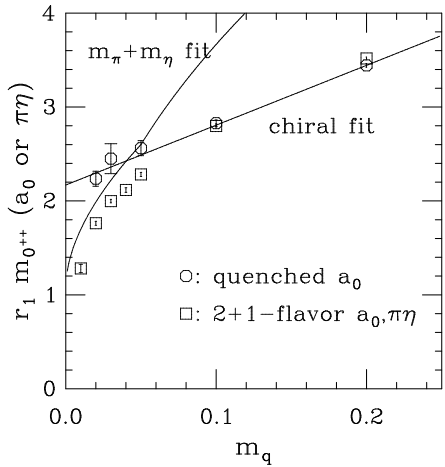,width=70mm}
 \vspace*{-11mm}
\caption{Mass of the $0^{++}$ state seen in our spectrum
analysis {\it vs} quark mass with and without dynamical
quarks.  The $\eta-\pi$ threshold is plotted.
\label{fig:a0_decay_clean} 
}
\vspace*{-8mm}
\end{figure}
As the pion mass drops to zero with decreasing quark mass, one expects
to see evidence for the decay of heavier mesons, such as $\rho
\rightarrow 2 \pi$ and $a_0 \rightarrow \eta \pi$.  The mixing between
a resonant state and a two-meson decay state should result in avoided
level crossing in the spectrum of the resonant state.  While such
decays are natural when dynamical quarks are present, quenched
simulations may also show hints of decay in some cases, through
``hairpin'' diagrams, although at different rates.

Previous attempts to observe effects of meson decay in the $\rho$
spectrum were unsuccessful \cite{Bernard:1993an}.  Here we report
results of a preliminary attempt to observe effects of the decay $a_0
\rightarrow \eta \pi$. This threshold is easier to approach than the
$\rho \rightarrow \pi+\pi$ threshold because the final pseudoscalars
don't have to carry away angular momentum, so they can both be at zero
momentum.  However, it is difficult, because with staggered quarks the
$a_0$ appears in the same propagators as the pions, so we must extract
the alternating exponential from a propagator containing a much larger
simple exponential.  We measure the $a_0$ mass ({\it i.e.} lowest
energy in the $a_0$ ($0^{++}$) channel) as a function of quark mass.
This is done on quenched and dynamical fermion configurations.
Preliminary results, shown in Fig.~\ref{fig:a0_decay_clean}, suggest
that the lowest $0^{++}$ mass drops more rapidly with decreasing quark
mass in the dynamical fermion simulation.  The drop appears to follow
the threshold energy of the decay channel, also shown.  Such a result
would be expected from avoided level crossing.  In fact, for the
lighter quark masses it is very difficult to extract masses for the
quenched $a_0$.  This may be a signal of the non-unitarized (because
of the quenching) couplings to two particle states.

\section{SUMMARY}

We continue studies using a recently proposed improved staggered
fermion action.  We find that the light hadron spectrum in
our currently preferred Asqtad action scales very well.  We have begun
a systematic study of quark loop effects with $2+1$ flavors.  Our
present sample of gauge configurations has been constructed so that
the lattice spacing is kept constant as the quark masses are varied.
We see no quark loop effects in the light hadron spectrum with $2+1$
flavors at a lattice spacing of $a = 0.14$ fm.  However, sea quarks
evidently modify the heavy quark potential both at short and long
distance, suggesting caution in using this potential to set a
precision lattice scale at unphysical quark masses.  Preliminary
determinations of the mass of the $a_0$ as a function of quark mass
suggest evidence for the decay to $\eta + \pi$.

This work is supported by the US National Science Foundation and
Department of Energy and used computer resources at Los Alamos National
Lab, NERSC, NCSA, NPACI
and the University of Utah (CHPC).

\end{document}